\documentclass[journal=jacsat,manuscript=article]{achemso}

\usepackage[version=3]{mhchem} 
\usepackage{graphicx}%
\usepackage{multirow}%
\usepackage{amsmath,amssymb,amsfonts}%
\usepackage{amsthm}%
\usepackage{mathrsfs}%
\usepackage[title]{appendix}%
\usepackage{xcolor}%
\usepackage{textcomp}%
\usepackage{manyfoot}%
\usepackage{booktabs}%
\usepackage{algorithm}%
\usepackage{algorithmicx}%
\usepackage{algpseudocode}%
\usepackage{listings}%

\author{Elisabeth Gruber}
\author{Lars H. Andersen}
\email{lha@phys.au.dk}
\affiliation[AU]{Department of Physics and Astronomy, Aarhus University, 8000 Aarhus C, Denmark}
\author{Laurence H. Stanley}
\author{Jan R. R. Verlet}
\email{j.r.r.verlet@durham.ac.uk}
\affiliation[DU]{Department of Chemistry, Durham University, Durham DH1 3LE, United Kingdom}
\author{Ivan S. Avdonin}
\author{Anastasia V. Bochenkova}
\email{bochenkova@phys.chem.msu.ru}
\affiliation[MSU]{Department of Chemistry, Lomonosov Moscow State University,  Moscow 119991, Russia}

\title[Dark-State-Mediated Efficient Energy Trapping in a Model GFP Chromophore]{Dark-State-Mediated Efficient Energy Trapping in a Model GFP Chromophore}


\begin{document}
\def\0S{S$_0$}
\def\1S{S$_1$}
\def\2S{S$_2$}
\def\nS{S$_n$ }
\def\ca{$\sim$}
\newcommand{\pil}{$\rightarrow$}
\newcommand{\plusminus}{$\pm$}

\begin{tocentry}
\includegraphics[]{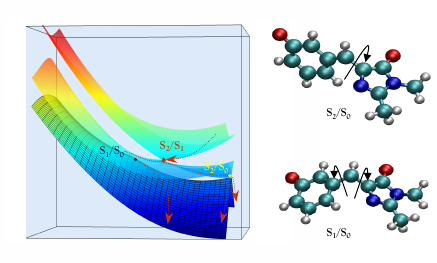}
\end{tocentry}

\begin{abstract}
  The functional properties of photoactive proteins are governed by the interplay between bright and dark excited states. While the bright states are well-studied, the dark states, which are fundamental to photostability and light harvesting, are notoriously difficult to characterize. Here, we report the direct observation and full characterization of an optically dark, low-lying singlet excited state in the isolated anion of the \textit{meta} green fluorescent protein (GFP) chromophore. Using a combination of ultrafast time-resolved action-absorption and photoelectron spectroscopy, we have captured the formation of this state in 100 fs and measured its remarkably long lifetime of 94 ps. We unambiguously assign its charge-transfer character and reveal the precise trapping mechanism through high-level \textit{ab initio} calculations. Our findings uncover a photoprotective mechanism in biomolecular anions where ultrafast internal conversion quenches electron emission, stabilizing long-lived electronic excitation even when the energy exceeds the electron detachment threshold.
\end{abstract}

\section{Introduction}
At first sight, one might think that optically dark states ({\it i.e.} electronic states having a small dipole coupling with the electronic ground state, or dark because of symmetry reasons) are of little interest. They do, however, play an important role in the photophysics of chromophores as they may be populated through conical intersections after photoexcitation to higher lying bright electronic states. Upon internal conversion, they may thus act as `trapping states' with long lifetimes and little fluorescence.
In photosynthetic complexes, symmetry-forbidden dark states are relevant in light-harvesting and for photoprotection  by carotenoids~\cite{Sundstrom-2004,Scholes2011,Grondelle2012}. When excited to the bright \2S state, the carotenoid chromophore relaxes within 50-300 fs to the optically inactive \1S state, which is then involved in the pigment-pigment energy transfer process. In carotenoids containing carbonyls, it is found that the lifetime of the dark charge-transfer states is strongly solvent dependent~\cite{Frank-2000}. 

Chromophores featuring long-lived dark excited states are of considerable interest for their potential in advanced technologies. They might be applicable in biomimetic electronic and memory devices, where there is a need to switch between light emitting and non-emitting states~\cite{WISE-2002}. 
The utility of these states also extends to sensing and photoprotection. Their long lifetime may enable their use as ''dark donors`` in F{\"o}rster resonance energy transfer, providing a background-free signal for biosensing~\cite{FRET2019}. 
Simultaneously, their capacity to absorb harmful high-energy photons via a higher-lying bright state and to dissipate the energy as heat from a stable dark state makes them highly efficient energy sinks, offering a strategy for enhancing the durability of polymers and synthetic materials~\cite{Yousif2013}. Given the above, it is not surprising that there has been a desire to develop photo-active proteins with engineered dark states. Specifically, dark charge-transfer states have been studied in the chromophores of the photoactive yellow protein (PYP) \cite{Rocha-Rinza-2010} and the Green Fluorescent Protein (GFP) \cite{Dong-2007,solntsev-2008,lha2014AngChem} through their meta-derivatives.

Solvation can strongly influence dark states, underscoring the need to understand their intrinsic spectroscopy and dynamics. Gas-phase spectroscopy uniquely enables such studies by isolating the intrinsic properties. Yet, the optically forbidden nature of dark states renders their investigation experimentally challenging, especially in the gas phase where ion densities are low. To overcome this problem we apply action-absorption spectroscopy, developed for studies of chromophore ions at ion-storage rings both in terms of spectroscopy~\cite{nielsen2001,Andersen_2004} and lifetimes~\cite{Svendsen_2017,Kiefer_2019}, alongside photoelectron spectroscopy~\cite{Verlet2014_PES,Anstoter2016} and  high-level quantum chemical calculations~\cite{BOCHENKOVA2024141,bochenkova2013ultrafast,Bochenkova2017}.

Here we focus on the chromophore of the GFP. This protein, with 238 amino acids, has a rigid 11-stranded $\beta$-barrel structure with a 4-(p-hydroxybenzylidene)-5-imidazolinone (pHBI) chromophore at the centre, well protected from external solvents. The protein is widely used for imaging purposes in biological samples due to its fluorescent properties~\cite{Tsien1998AnnurevBiochem}. Numerous studies of GFP-model chromophores have been conducted over the past almost twenty years. In particular, a dimethyl derivative of pHBI, called para-HBDI is used as a model for the chromophore of the protein. For this molecule, the absorption spectrum within the visible range is dominated by the intense \0S$\rightarrow$\1S transition at $\sim$480~nm~\cite{nielsen2001} (see Fig.~\ref{fig:absorption}), with a band origin, determined by cryogenic spectroscopy, at 481.5~nm~\cite{Andersen2023}. Above the electron-detachment threshold, electron emission proceeds through autodetachment of vibrational resonances (VRs), whereas below the threshold there may appear a significant contribution from sequential multiphoton absorption in action-absorption spectra mediated by the strong \0S$\rightarrow$\1S transition in para-HBDI \cite{nielsen2001,toker2012direct,bochenkova2013ultrafast,lha2014AngChem}. The separation between one and multiphoton contributions is conveniently done by the use of fast fs-laser pulses \cite{Hjalte_PRL2016}. 

Moving the oxygen atom on the phenolate ring from the {\it para} to the {\it meta} position (see Fig.~\ref{fig:absorption}) significantly alters the low-energy part of the HBDI photoabsorption spectrum, which is relevant to its photochemistry~\cite{lha2014AngChem}. The absence of valence resonance structures of the meta-HBDI anion in the electronic ground state causes an effective decoupling of the isoelectronic $\pi$-systems of the phenolate and imidazolinone rings. Calculations of the relevant $\pi$ and $\pi^*$ orbitals reveal that the \0S$\rightarrow$\1S transition has a significant charge-transfer character, implying that this \1S state is strongly influenced by the solvation of a polar solvent ({\it e.g.} water). It is predicted to lie at the far red edge of the visible spectrum ($\sim$700 nm) with a very low oscillator strength~\cite{lha2014AngChem}.

This work offers direct spectroscopic access to the dark \1S state of meta-HBDI, revealing its lifetime and energy. The photoresponse of the anion has been studied in the spectral region from 350 to 700~nm in the gas phase. The presence of an optically dark, low-lying excited state, which has been predicted theoretically, is confirmed through action-absorption and photoelectron spectroscopy. The charge-transfer character of the \0S $\rightarrow$ \1S transition is affirmed by measuring a significant blue shift when the chromophore is complexed with betaine, a molecule with a dipole moment of almost 12~D. Excited-state dynamics is studied using both pump-probe action and photoelectron spectroscopies and is supported by {\it ab initio} calculations to provide a comprehensive understanding of the energy flow. We find a sub-nanosecond excited-state lifetime for the dark \1S state, whereas the bright \2S state decays to \1S on an ultrafast timescale of 100~fs. We show that internal conversion between S$_2$ and S$_1$ through a conical intersection dominates over electron detachment when the anion is excited at 400~nm, despite being well above its photodetachment threshold ($\sim$490~nm). This results in efficient trapping of electronic excitation in the dark excited state located in the near-IR region. Our results offer a new and exquisite window into the sensitivity and role of dark states in fluorescent proteins.

\section{Results and discussion}
\subsection*{Nature and Energy Landscape of Excited States}
\begin{center} 
\begin{figure}[!t]
\centering
\includegraphics[width=0.7\textwidth]{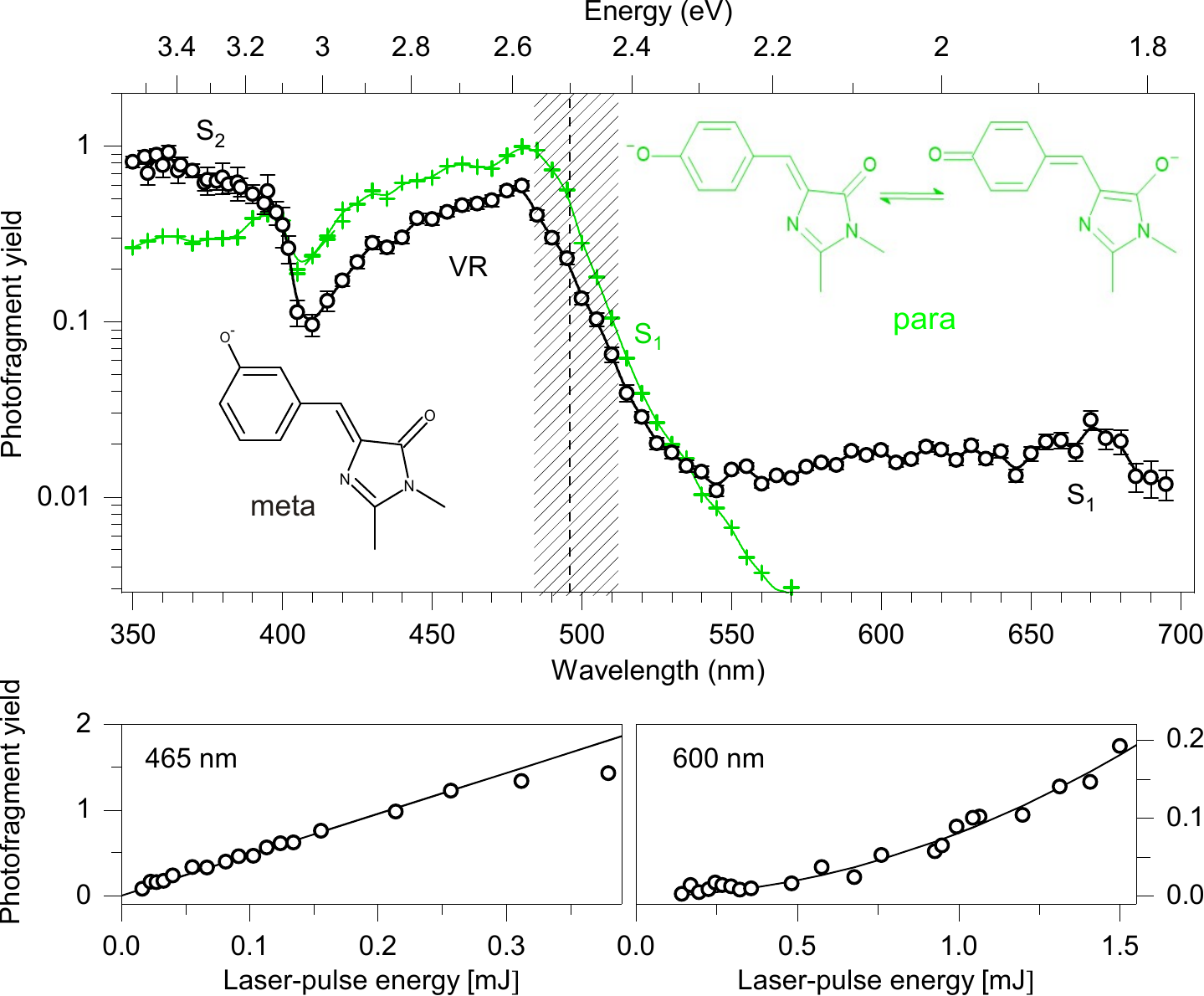}
\caption{Action-absorption spectrum of meta-HBDI (black data points) in comparison to para-HBDI (green data points). Notice the logarithmic vertical axis. The vertical line shows the electron-detachment threshold for the meta chromophore. Between 350 and 500~nm prompt action is detected after single-photon absorption (photofragment yield increases linearly with the laser-pulse energy, presented here for 465~nm in the lower graph) and associated with excitation into \2S and vibrationally resonant (VR) photodetachment out of the dipole-bound state. The weak absorption between 550 and 700~nm for meta-HBDI is attributed to excitation of the dark \1S  state. Sequential absorption of two photons is necessary to cause fragmentation (photofragment yield increases quadratically with the laser-pulse energy, presented here for 600~nm in the lower graph). Data in the region 550 and 700~nm is normalized according to the quadratic laser-power dependence.} 
\label{fig:absorption}
\end{figure}
\end{center}

Figure~\ref{fig:absorption} shows the absorption spectrum of meta-HBDI in the spectral region from 350 to 700~nm, covering transitions from the electronic ground state \0S into the first excited state \1S (weak absorption between 550~nm and 700~nm) and into the second excited state \2S (below 400~nm). The vertical electron detachment energy (VDE) for meta-HBDI was previously determined to be in the range of 2.40~--~2.54~eV $\pm$ 0.1~eV~\cite{lha2014AngChem}. The significant spread was due to inhomogeneous broadening caused by internal rotation about the single C-C bridge bond in the electronic ground state. Indeed, MP2/(aug)-cc-pVTZ calculations reveal two rotamers in meta-HBDI, separated by an energy of 0.05~eV with an interconversion barrier of 0.46~eV (see Fig.~S1). While these rotamers are spectroscopically indistinguishable at their equilibrium planar geometries, rotation along the interconversion coordinate lowers the VDE, with a maximum reduction of 0.2 eV at the transition state. Here, using photoelectron spectroscopy, we refine the experimental VDE and also find the adiabatic detachment energy to be 2.63 and 2.30 $\pm$ 0.05 eV, respectively.

The broad and intense absorption observed between 400 and 500 nm -- near and above the electron-detachment threshold -- arises from two main processes: direct electron detachment, and excitation of vibrational resonances associated with a non-valence, electronically dipole-bound state.~\cite{lha2014AngChem}
States with little vibrational excitation are truly bound and are observed experimentally through sequential two-photon absorption. In contrast, excess vibrational energy in the radical core can promote these states into the continuum, where they manifest as autodetaching resonances~\cite{Anstoter2016,Verlet2020,Andersen_DBS_2025}.
In meta-HBDI, direct transitions from \0S to these states are spectrally distinct from transitions to the valence excited states: \1S in the near-IR, which is electronically bound, and \2S in the UV, which is embedded in the continuum and is therefore metastable against electron emission.

\begin{center} 
\begin{figure}[!t]
\centering
\includegraphics[width=0.7\textwidth]{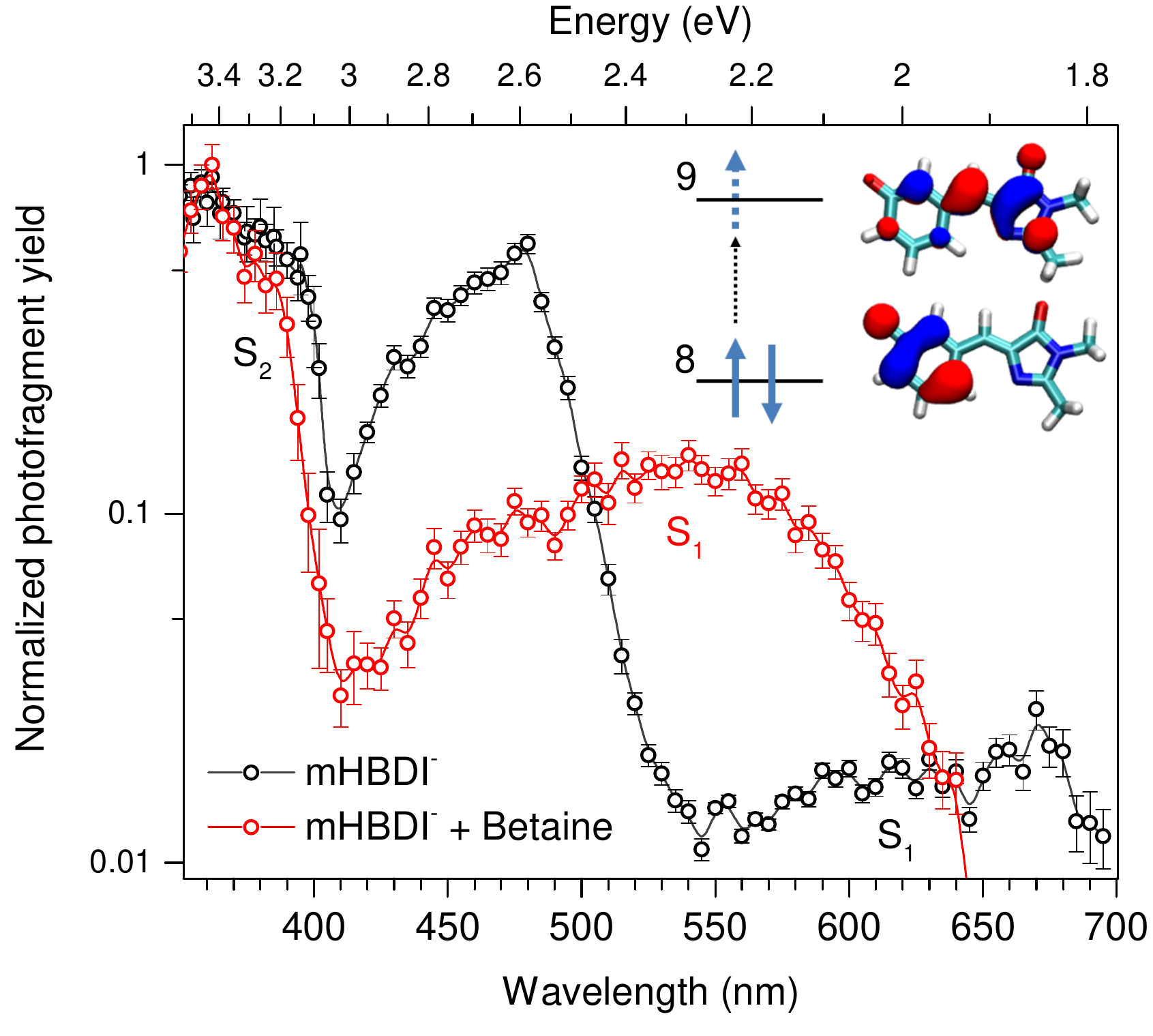}
\caption{Prompt action spectrum of meta-HBDI (black data points) in comparison with meta-HBDI complexed with betaine (red data points). A strong blueshift of the \0S-\1S transition energy in the meta-HBDI - betaine complex reflects a strong CT character. Calculations of the $\pi$ and $\pi$* orbitals, which are primarily involved in the excitation, show that the transition is connected to charge transfer from the phenolate ring into the imidazolinone ring. Only little blueshift for the transition into \2S is observed.} 
\label{fig:CT}
\end{figure}
\end{center}

The charge-transfer nature of the \0S $\rightarrow$ \1S transition was revealed by zwitterion tag action spectroscopy (ZITA spectroscopy)~\cite{stockett2017}. The excitation of a charge-transfer (CT) transition significantly redistributes electron density within the chromophore. This redistribution alters the Coulombic interaction with any nearby dipole, making CT transitions exceptionally sensitive to their electrostatic environment, including solvent interactions.
The different interaction strength between the ion and the dipole in the ground and excited states leads to a blueshift of the absorption band. Thereby, the extent of the shift is a measure of the CT degree. With a 11.9 D dipole moment arising from its quaternary ammonium and carboxylate termini \cite{shikata2002dielectric}, the zwitterion betaine is an excellent probe molecule for ZITA spectroscopy \cite{stockett2017,toker2018counterion,langeland2018}.

The absorption of the betaine-meta-HBDI complex shows a rather different picture with a more than 100~nm blue-shift for the \0S $\rightarrow$ \1S transition energy, pointing to a strong CT character (see Fig.~\ref{fig:CT}).
Indeed, the calculated $\pi$ and $\pi^*$ orbitals, which are primarily involved in the excitation, show that the \0S $\rightarrow$ \1S transition is associated with a significant electron-density transfer from the phenolate ring to the imidazolinone ring (see inset in Fig.~\ref{fig:CT}). Only a small blueshift is observed for the 400 nm transition into \2S, pointing to a much lower CT character. 

\begin{center}
\begin{figure}[!t]
\centering
\includegraphics[width=0.6\textwidth]{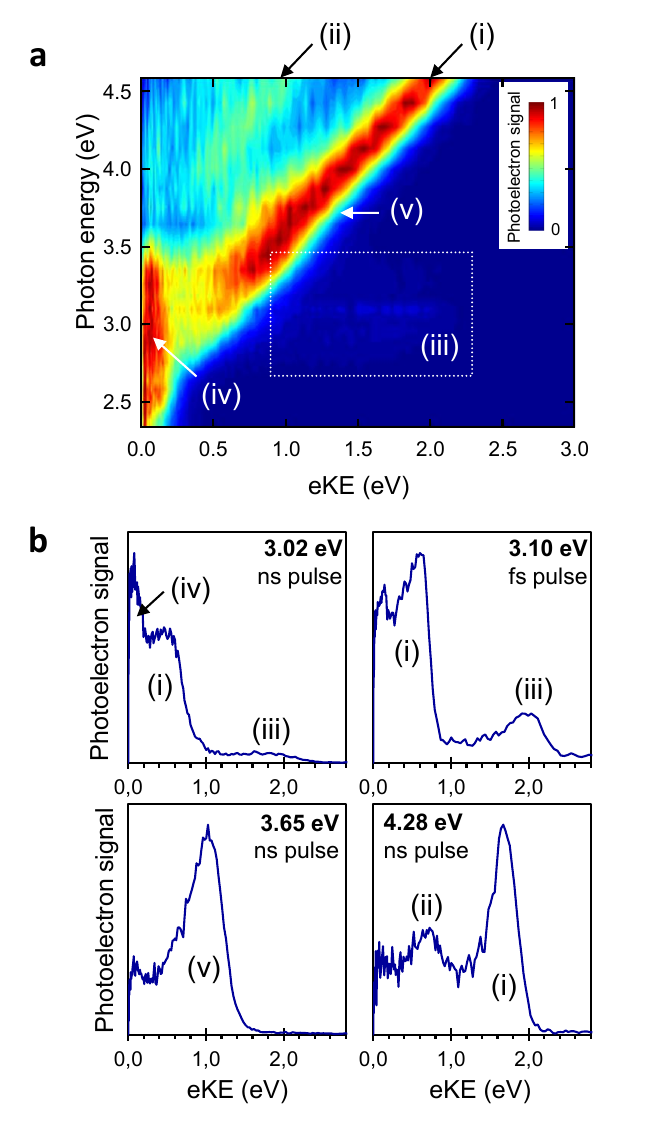}
\caption{Photoelectron spectra of the meta-HBDI anion. (a) Two-dimensional photoelectron spectrum recorded with nanosecond laser pulses. (b) One-dimensional spectral cuts at specified excitation energies. For comparison, the spectrum at 3.10~eV was acquired with femtosecond pulses, contrasting with the 3.02~eV spectrum measured using ns pulses.}   
\label{fig:ExPES}
\end{figure}
\end{center}

\begin{figure}[!t]
\centering
\includegraphics[width=0.9\textwidth]{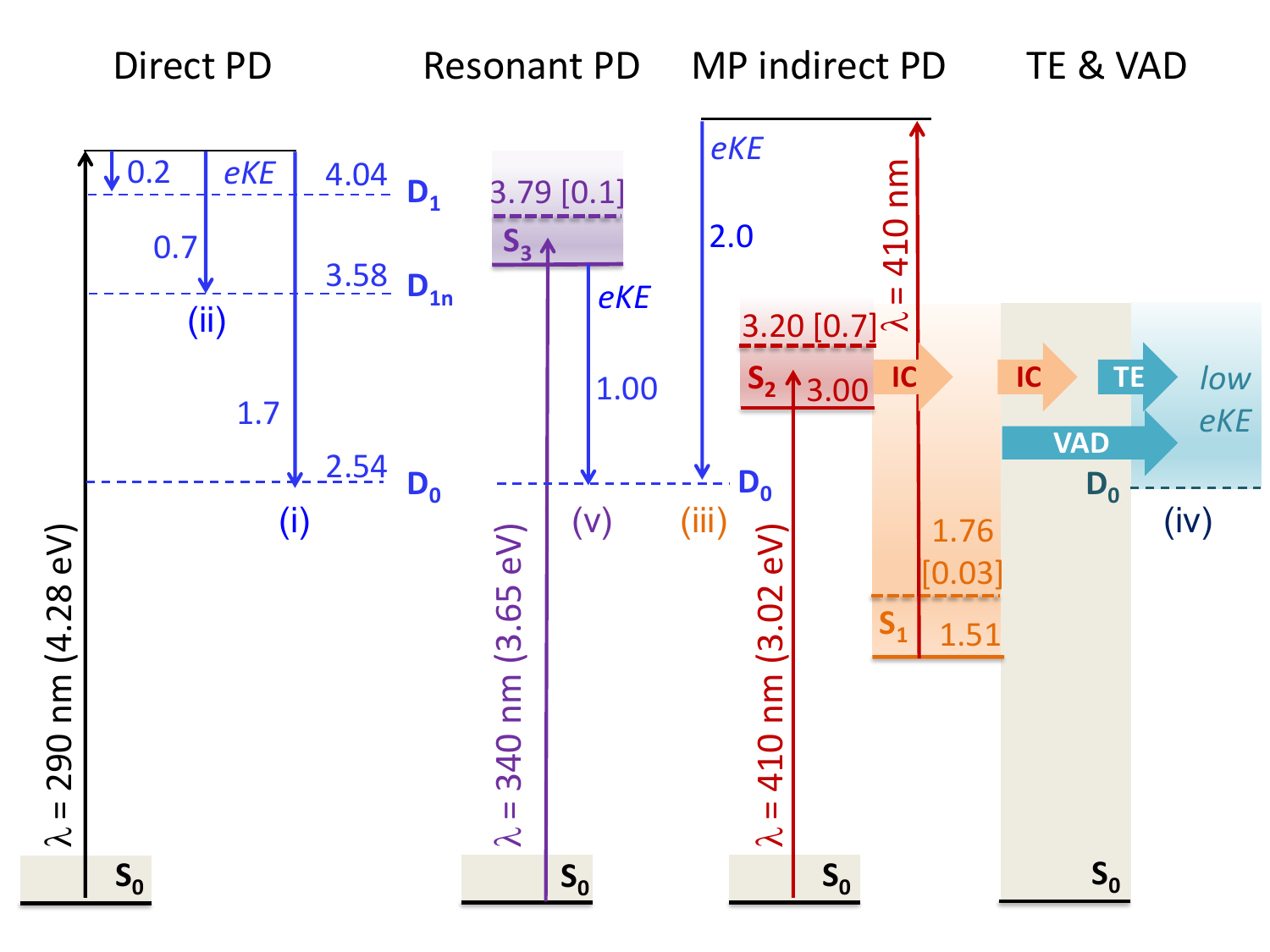}
\caption{Wavelength-dependent photodetachment (PD) mechanisms for the meta-HBDI anion. The energy-level diagram, calculated at the XMCQDPT2/SA(3)-CASSCF(16,14)/(aug)-cc-pVDZ level, shows pathways for different excitation energies: direct PD; resonant PD through the S$_3$ shape resonance; multiphoton (MP) indirect PD via the \1S state after internal conversion (IC) from the \2S Feshbach resonance; thermionic emission (TE) from \0S; and vibrational autodetachment (VAD) from \1S. Channels ({\it i}) -- ({\it v}) are identified in the experimental 2D photoelectron spectrum (Fig.~\ref{fig:ExPES}). Energies (in eV) and oscillator strengths (in brackets) are provided for key transitions. Adiabatic and vertical energies are shown as solid and dashed lines, respectively, with colored areas representing the vibrational manifold. Blue arrows indicate the electron-detachment process, labeled with the resulting electron kinetic energy (eKE).} 
\label{fig:Diagram}
\end{figure}

The presence of the low-lying dark state in the meta-HBDI anion is further supported by photoelectron spectroscopy and high-level {\it ab initio} calculations. Figure~\ref{fig:ExPES} shows the two-dimensional photoelectron spectrum~\cite{Anstoter2016,Verlet2020_2D,Verlet2021_2D} of meta-HBDI recorded with nanosecond laser pulses with one-dimensional cuts at specific photon energies. Several distinct regions, labeled ({\it i}) -- ({\it v}), are highlighted for discussion. With reference to the energy levels, calculated using multiconfiguration quasi-degenerate perturbation theory XMCQDPT2~\cite{Granovsky2011} (see Fig.~\ref{fig:Diagram} and the SI for computational details), these regions correspond to the following processes: 

\begin{enumerate}
  \renewcommand{\labelenumi}{\roman{enumi})}
    \item The linear relation between the photon energy and the energy of the outgoing electron corresponds to direct photodetachment h{\it v} + S$_0$  $\rightarrow$  D$_0$ + e$^-$. Excitation at 4.28~eV yields the most unperturbed photoelectron spectrum (see Fig. \ref{fig:ExPES}b, lower right panel). We determine adiabatic and vertical electron detachment energies of 2.30 $\pm$ 0.05~eV and 2.63 $\pm$ 0.05~eV, respectively, which are consistent with and refine the previously measured values of 2.40 -- 2.54 $\pm$ 0.1~eV ~\cite{lha2014AngChem}. The calculated vertical detachment energy (D$_0$) is 2.54 eV, which corresponds well to the ridge of ({\it i}) in the photoelectron spectrum.  
    \item Similarly, region ({\it ii}) corresponds to direct detachment to an excited state in the neutral radical D$_{1n}$. The calculated difference between D$_{1n}$ and D$_0$ is 1.04~eV, which matches the experimental gap.    
    \item The appearance of region ({\it iii}) with an electron-kinetic energy higher than that of direct one-photon detachment provides evidence for a two-photon detachment process. This mechanism requires the first photon to populate an intermediate state of the anion, from which a second photon, absorbed within the same laser pulse, causes electron detachment. In the ns-laser experiments, the weak signal in region ({\it iii}) is assigned to photodetachment from the intermediate \1S state. Electrons ejected from this state are born with an adiabatic excitation energy (AEE) of 1.51 eV (see Fig.~\ref{fig:Diagram}). For a process where \1S is populated by internal conversion from \2S, the sequential absorption of two 3.02 eV photons yields photoelectrons with a maximum kinetic energy of AEE(S$_1$) + h$v$ – VDE(D$_0$) $\sim$~2.0 eV (Fig.~\ref{fig:ExPES}b, upper left panel). The peak is significantly enhanced with 400~nm (3.1 eV) fs-laser pulses (Fig.~\ref{fig:ExPES}b, upper right panel) due to the higher probability of absorbing a second photon from the intermediate \1S state within the tens-of-fs pulse duration compared to the much longer 5~ns pulse. Although absorption from \2S could yield a peak at AEE(S$_2$) + h$v$ – VDE(D$_1$) $\sim$ 2.0 eV, its lifetime is much shorter (see below) than that of \1S and comparable to the femtosecond pulse duration, rendering this pathway less probable.
    \item The region of low-energy electrons ({\it iv}) is due to internal conversion from \2S to \1S and further to \0S followed by statistical thermionic emission from the ground state~\cite{Cambell_1991} or vibrational autodetachment out of the electronically bound \1S state~\cite{Andersen2023}. This channel may also contain a contribution from sequential absorption of multiple photons from the ns-laser pulses (duration $\sim$5 ns). Internal conversion is here faster than autodetachment out of \2S. This resonance state is of a Feshbach type with respect to the open D$_0$ continuum~\cite{lha2014AngChem}, and hence electronic autodetachment out of this state is a two-electron process, which becomes slower than the sub-ps nuclear photoresponse in this region~\cite{Horke2013,Verlet2015_ChemSci}.
    \item Finally, in region ({\it v}) between the 3.6 eV and 3.9 eV photon energy there is a slight broadening in the signal, which may be caused by the opening of the resonant channel h{\it v} + S$_0$  $\rightarrow$ S$_3$ $\rightarrow$ D$_0$ + e$^-$ (Figs.~\ref{fig:ExPES}a and~\ref{fig:ExPES}b, lower left panel). The calculated position of S$_3$ supports this assignment (see Fig.~\ref{fig:Diagram}). This state is a shape resonance with respect to the D$_0$ continuum~\cite{lha2014AngChem}, and one-electron autodetachment dominates the signal in region ({\it v})~\cite{Verlet2015,Bochenkova2017}.
\end{enumerate}

We have shown that action-absorption spectroscopy has direct spectroscopic access to the near-IR dark \1S state of charge-transfer character. The presence of this electronically bound state is also confirmed by photoelectron spectroscopy. It is observed through a multi-photon indirect photodetachment process: initial excitation populates the bright \2S state above the detachment threshold, followed by internal conversion to the \1S state and subsequent electron detachment via absorption of a second photon. From the presence of the low-eKE part in the two-dimensional photoelectron spectrum, it also follows that internal conversion from \2S competes with electron autodetachment out of this resonance state. In the following, we present direct measurements of the \2S and \1S lifetimes using femtosecond pump–probe spectroscopy that integrates time-resolved action-absorption and photoelectron detection.

\subsection*{Dynamics of Excited States}

\begin{center} 
\begin{figure}[H]
\centering
\includegraphics[width=0.5\textwidth]{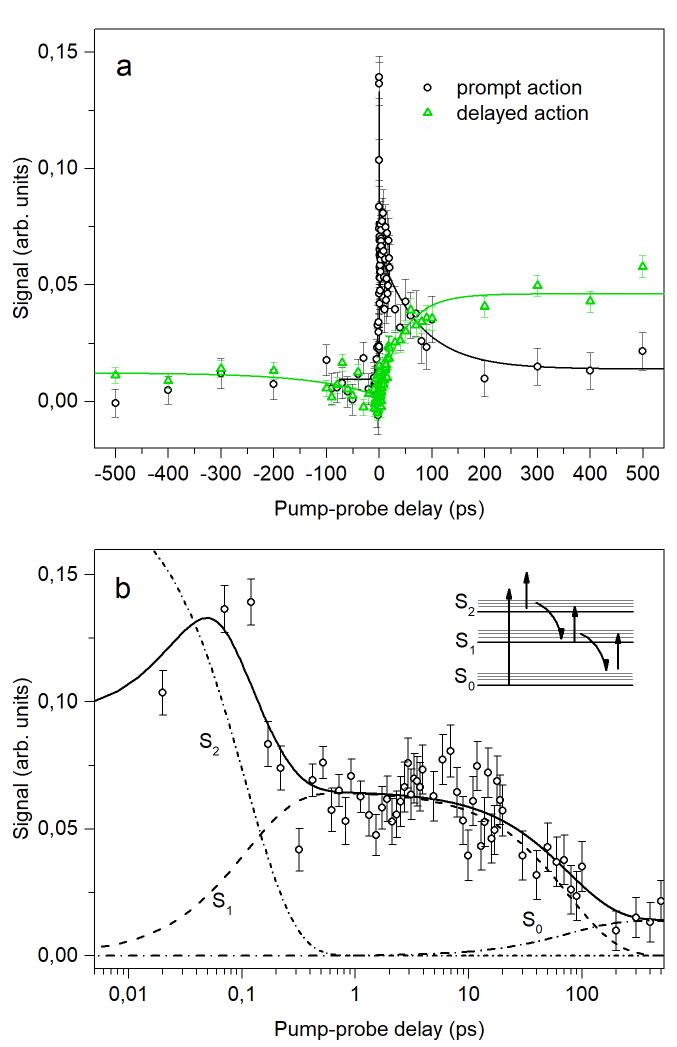}
\caption{Excited-state lifetime measured in pump-probe experiments at the ion storage ring SAPHIRA. The molecules were first excited into \2S by a 400~nm pump pulse and then probed by a 800~nm probe pulse. The experimental setup enables the detection of prompt (within the first 10~$\mu$s after photoexcitation) as well as delayed (after 10~$\mu$s up to several ms after photoexcitation) photofragmentation. The upper graph (a) shows the prompt and delayed fragmentation as a function of the pump-probe delay on a linear timescale. While the lifetime of \2S and \1S can be probed by detecting prompt fragmentation, the delayed action shows the ground state recovery. In the lower graph (b), the signal of the prompt action is plotted on a logarithmic time scale and shows that the decay consists of a fast and a slow component. The fast decay of 100~fs $\pm$ 26~fs corresponds to the fast relaxation out of \2S with subsequent trapping in \1S for 94~ps $\pm$ 11~ps. The solid line represents a fit through the data points using a rate model indicated by the inserted schematic. The dashed lines show the calculated population of \2S, \1S and \0S.} 
\label{fig:pump_probe}
\end{figure}
\end{center}

The femtosecond pump-probe technique combined with time-resolved action-absorption spectroscopy enables simultaneous monitoring of both excited-state decay and ground-state recovery. Figure~\ref{fig:pump_probe} shows the measured prompt and delayed photo-induced yields of neutrals. The fast $\sim$100~fs decay of the prompt signal is attributed to ultrafast internal conversion from S$_2$ to S$_1$. The slower 94~ps decay, appearing as the long-lived decaying component of the prompt response, is attributed to the decay out of \1S. It corresponds to the repopulation of the S$_0$ ground state, fully consistent with the recovery dynamics revealed in the delayed action.

\begin{center} 
\begin{figure}[!b]
\centering
\includegraphics[width=0.6\textwidth]{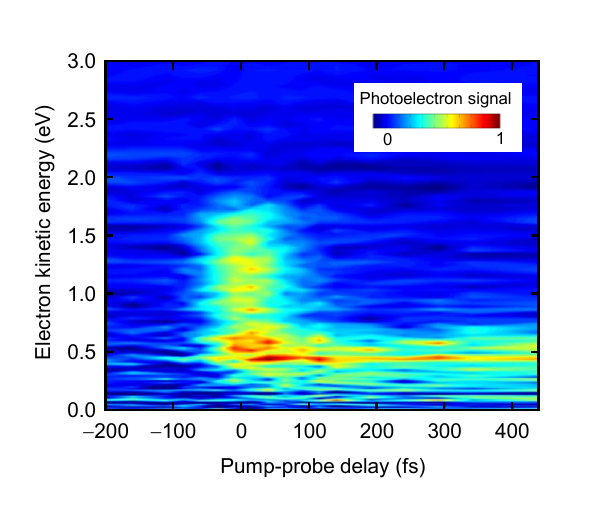}
\caption{Photoelectron yield in pump-probe measurements (400~nm + 800~nm). The signal between 0.5 and 1.5~eV decays rapidly and is linked to the relaxation out of the \2S state. The signal at 0.4~eV remains for a longer time due to population trapping in the dark \1S state from the \2S state. The ground-state recovery is not detected due to insufficient photon energy in the 800 nm probe pulse.} 
\label{fig:electron_spectroscopy}
\end{figure}
\end{center}

The time-resolved photoelectron spectrum is plotted in Fig.~\ref{fig:electron_spectroscopy}. It is evident that also here a very fast as well as a long-lived component exist. The fast component, which has a kinetic energy extending to $\sim$1.7~eV also decays within the first 100~fs. The spectrum of this feature is consistent with detachment from the \2S state. The \2S-signal decay is commensurate with the appearance of a new spectral feature with a kinetic energy peaking around 0.45~eV, corresponding to a binding energy of about 1.1~eV. This can be assigned to the \1S-excited state, which lies about 1.2~eV lower than the \2S state, according to Fig.~\ref{fig:absorption}. As soon as the electronic ground state is reached, the absorption of a 1.55~eV (800~nm) probe photon does not suffice to lead to electron detachment. The \2S excited-state signal is observed to decay with a lifetime of $\sim$90 fs, which is in excellent agreement with the measurement by the action-absorption spectroscopy technique.

\subsection*{Trapping Mechanism}

The experimental results are rationalized by the aid of {\it ab initio} calculations. We find that the S$_1$ state of the meta-HBDI anion is optically dark in the Franck-Condon region, whereas the \0S $\rightarrow$ \2S transition is bright (see Fig.~\ref{fig:Diagram}). The calculated vertical excitation energy of the \0S $\rightarrow$ \1S transition is 1.76~eV (704~nm) with an oscillator strength of 0.03. The red-shifted value and the very low intensity of the first excitation are attributed to its nature connected to the transfer of charge and electron density from phenolate to the imidazolinone ring (see Figs. S2-S3). The second transition is the brightest in the absorption spectrum, with a calculated vertical excitation energy of 3.2 eV (387~nm) and an oscillator strength of 0.7. It involves a local redistribution of electron density within the $\pi$-system of the imidazolinone ring (Figs. S2-S3).

In the present study, ZITA action-absorption spectroscopy of the meta-HBDI anion complexed with betaine directly confirms the high charge-transfer character of the \1S state and the low charge-transfer character of the \2S state. The energetics of the \1S state is also confirmed experimentally. In action-absorption spectroscopy, a broad absorption profile between 550–700~nm is assigned to direct excitation of \1S. The broad Franck-Condon envelope of this band is characteristic of a charge-transfer transition. Furthermore, photoelectron spectroscopy gives a vertical electron binding energy of 1.1~eV for the \1S state. When combined with the measured VDE(D$_0$) of 2.6~eV, this yields an \0S $\rightarrow$ \1S adiabatic excitation energy of 1.5~eV, a value in excellent agreement with the calculated result of 1.51~eV.

\begin{center} 
\begin{figure}[t!]
\centering
\includegraphics[width=0.7\textwidth]{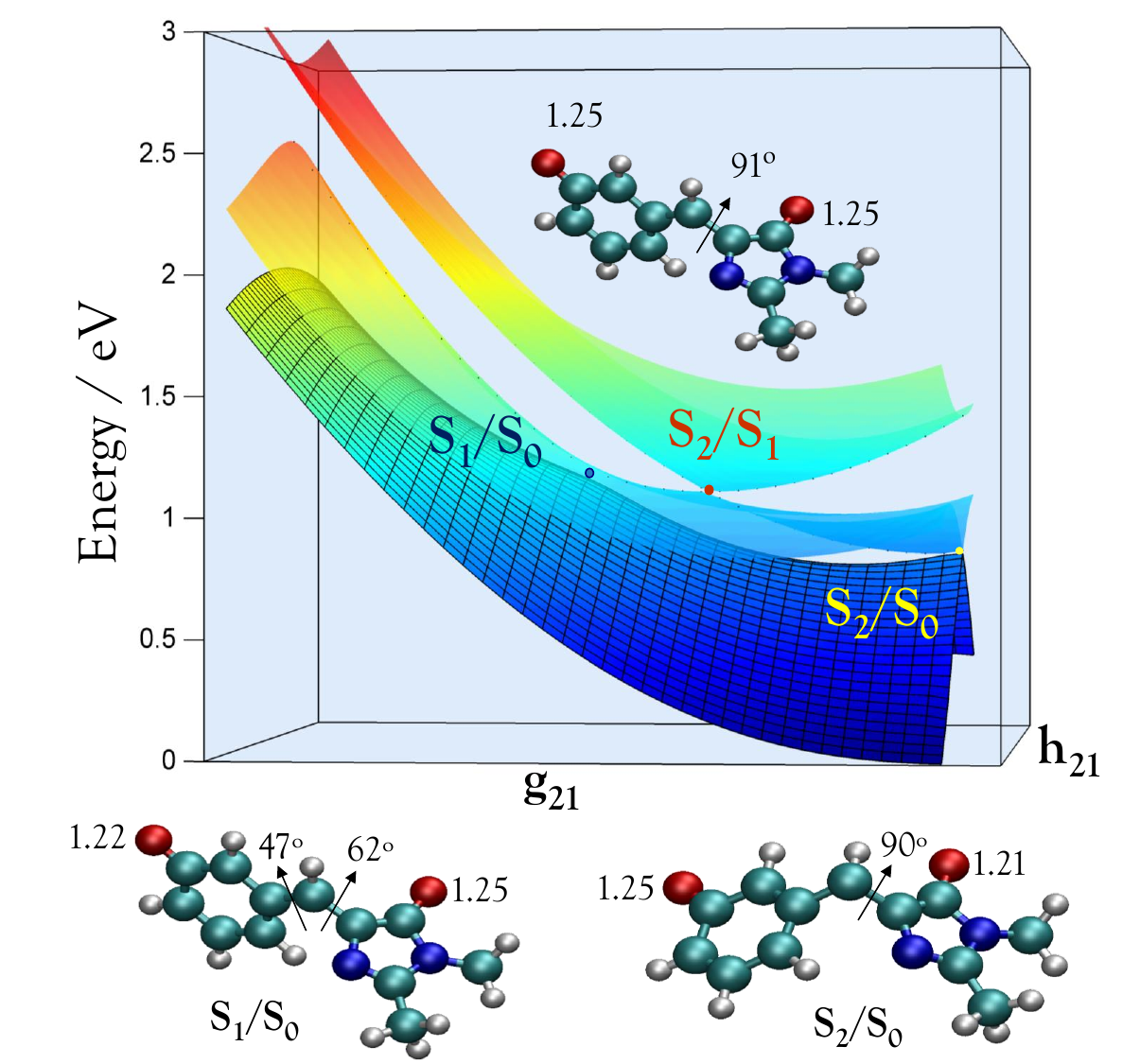}
\caption{CASSCF(16,14)/(aug)-cc-pVDZ potential-energy surface illustrating the conical intersections governing the excited-state dynamics of the meta-HBDI anion. The surface is plotted in the branching plane of the  S$_2$/S$_1$ conical intersection, spanned by the gradient difference vector (\textbf{g$_{21}$}) and the nonadiabatic coupling vector (\textbf{h$_{21}$}). Also shown are the minimum-energy conical intersection structures, with key torsional angles in the bridge moiety and O–C bond lengths (in \AA) indicated. This representation highlights how nuclear motion along these directions defines the nonradiative decay along two competing branches involving three conical intersections. Note that the intermediate \1S state traps the excited-state population at the planar equilibrium geometry. The \1S/\0S minimum-energy conical intersection on the dark branch is the lowest-energy pathway for internal conversion to the ground state (see Fig.~\ref{fig:Pump_Probe_Diagram}).}  
\label{fig:PES}
\end{figure}
\end{center}

Analysis of electron density and charge distribution in \2S and \1S reveals that excitation dramatically reduces the $\pi$-bond order of a key C=C bridge bond -- by 2.2 in \2S and 1.6 in \1S -- which equalizes the bridge bonds and facilitates ultrafast internal conversion via bond rotation. Furthermore, excited-state resonance couples the valence structures of \2S and \1S, scrambling their isoelectronic $\pi$-subsystems and equalizing charge localization probability between the phenolate oxygen and the bridge carbon, thereby altering the effective $\pi$-system partition (see Fig.~S4). Rotation around the bridge C=C bond serves as the key reaction coordinate, driving the system toward the S$_2$/S$_1$ conical intersection by equalizing charge localization probability between the states. This leads to the highly efficient internal conversion observed between the \2S and \1S states.

Three conical intersections (CI) are located that interconnect the S$_2$, S$_1$, and S$_0$ states, which are relevant to the relaxation dynamics of the meta-chromophore in the gas phase. Fig.~\ref{fig:PES} shows the potential energy surfaces of S$_2$, S$_1$, and S$_0$ in the branching plane of the \2S/\1S CI. The initial relaxation through the peaked \2S/\1S CI occurs barrierless (Fig.~S5) and is therefore very fast, in agreement with the experimental finding. The decay results in population of the \1S state despite the open electron-detachment channel.
However, a bifurcation at the \2S/\1S CI funnels the \1S population toward two distinct conical intersections with the ground state (Fig.~\ref{fig:PES}). These two CIs are differentiated by the character of the first excited state: one acquires the CT character of the dark \1S state, while the other retains the character of the bright \2S state. Therefore, they are labeled as \1S/\0S and \2S/\0S, respectively.

\begin{figure}[!t] 
\centering
\includegraphics[width=0.9\textwidth]{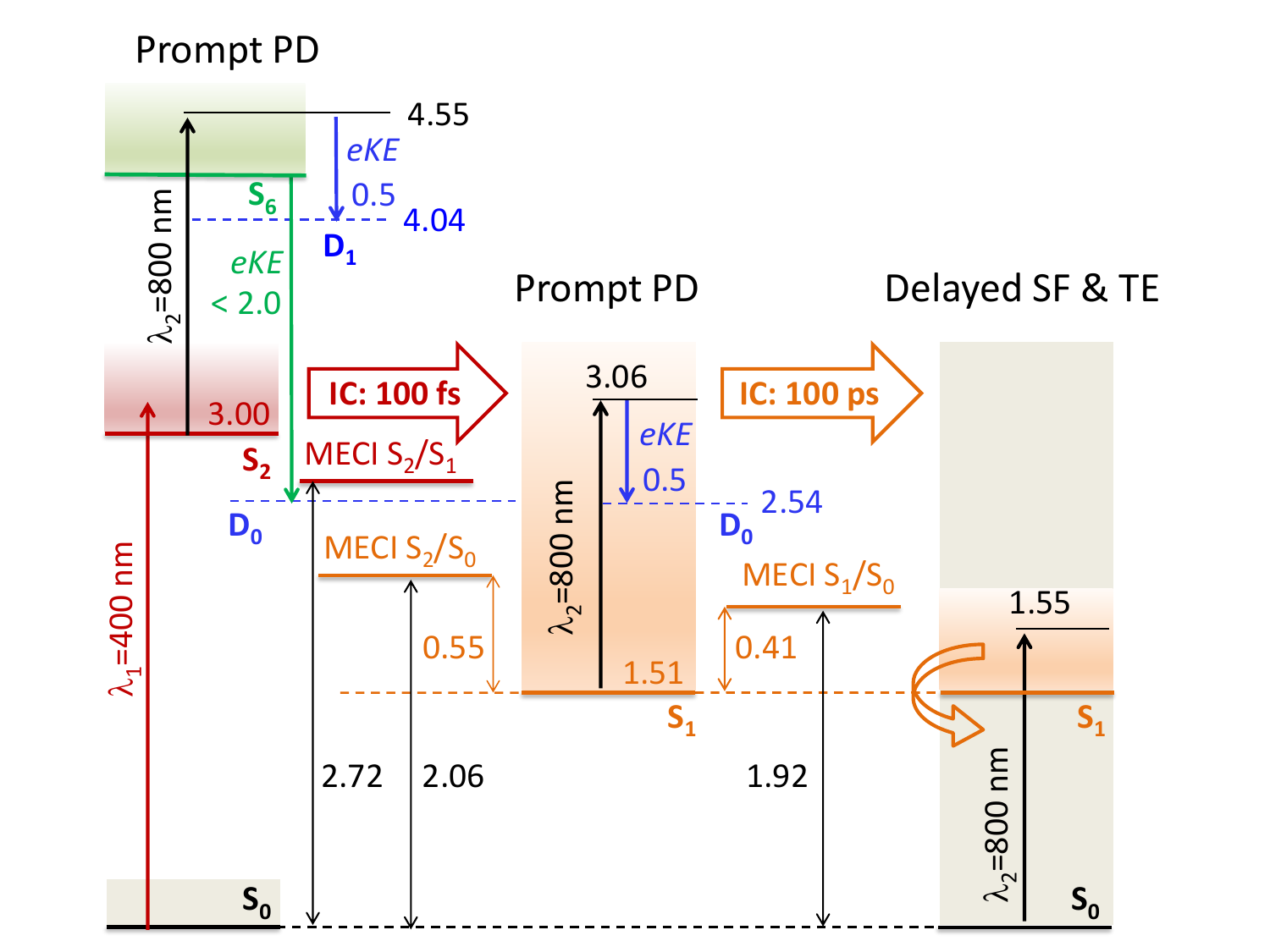}
\caption{Trapping mechanism in the meta-HBDI anion. The XMCQDPT2/SA(3)-CASSCF(16,14)/(aug)-cc-pVDZ diagram shows relative energies of key electronic states and minimum-energy conical intersections (MECI) between the \2S, \1S, and \0S states. Shown are also the prompt and delayed channels in pump ($\lambda_1$) -- probe ($\lambda_2$) spectroscopy. The excited-state decay is registered both by measuring the energy of emitted electrons and by counting neutrals in prompt photodetachment (PD) channels. The ground-state recovery is measured as a delayed signal through counting neutrals following statistical fragmentation (SF) and thermionic emission (TE) from \0S. Energies are in eV.}
\label{fig:Pump_Probe_Diagram}
\end{figure}

The calculated energy-level diagram identifies the location of the minimum-energy conical intersections (MECIs) governing the non-adiabatic transitions between the \2S, \1S, and \0S states (see Fig.~\ref{fig:Pump_Probe_Diagram}). Following efficient internal conversion from \2S, the meta-HBDI anion becomes trapped in the planar equilibrium geometry of the \1S state. This trapping occurs because the \2S/\0S and \1S/\0S MECIs lie 0.55~eV and 0.41~eV higher in energy, respectively. The two decay pathways are structurally distinct: the \2S/\0S CI involves rotation about the double C=C bridge bond, a pathway that could lead to isomerization. In contrast, the \1S/\0S CI is characterized by twisting about both bridge bonds and significant pyramidalization of the bridge carbon (see Fig.~\ref{fig:PES} and Fig.~S6). In the gas phase, the relaxation proceeds predominantly along this latter pathway, resulting in a long-lived non-fluorescent trapping state that ultimately decays to the ground state without isomerization.

The photoinduced dynamics, studied using femtosecond pump–probe spectroscopy, is characterized by two complementary methods: (1) time-resolved action-absorption spectroscopy, which monitors prompt and delayed neutrals; and (2) photoelectron spectroscopy, which detects prompt electrons generated by direct photodetachment from the excited states. Direct detachment out of \2S should comply with the Slater rules. In other words, two-electron processes are forbidden in the dipole approximation. Therefore, following 400~nm excitation of \2S in meta-HBDI, an 800~nm probe photon can detach an electron from \2S with the most probable eKE of 0.5~eV (D$_1$), leaving the neutral with 0.1~eV of vibrational energy (Fig.~\ref{fig:Pump_Probe_Diagram}). {Alternatively, the probe photon can promote the anion from \2S to a higher-lying resonance at $\sim$4.6 eV, which autodecays, producing a broad electron kinetic energy distribution up to 2.0~eV across all open continua (D$_1$, D$_{1n}$, and D$_0$). The experimental eKE distribution of up to $\sim$1.7~eV observed at early times (Fig.~\ref{fig:electron_spectroscopy}) is consistent with both direct and resonant photodetachment pathways, which leave some vibrational energy in the neutral core. This is supported by the XMCQDPT2 calculations, which identify an excited state at 4.6~eV at the ground-state equilibrium geometry, with a non-negligible oscillator strength of 0.06 for the \2S $\rightarrow$ S$_6$ transition (see Fig.~\ref{fig:Pump_Probe_Diagram}).

If internal conversion to \1S occurs first, the same probe detaches an electron from \1S, producing electrons with an energy of 0.5~eV (D$_0$) and a vibrationally hot neutral with 1.59~eV of excess energy. The significant narrowing of the eKE distribution peaking at 0.45~eV after 100~fs (Fig.~\ref{fig:electron_spectroscopy}) confirms the direct photodetachment pathway. Repopulation of the \2S state from \1S by the 800~nm probe is negligible due to the vanishing oscillator strength of this transition (Fig.~\ref{fig:Pump_Probe_Diagram}). Following internal conversion to \0S, the 800~nm excitation leads to repopulation of the electronically bound \1S state, resulting in slow SF and TE from the hot ground state after internal conversion. The excited-state decay and ground-state recovery align well with the prompt and delayed signals observed in our pump-probe action-absorption experiments (Fig.~\ref{fig:pump_probe}).

The excited-state population is trapped in \1S due to the presence of a barrier along the minimum-energy pathway that leads from the planar S$_1$ equilibrium structure to the highly twisted S$_1$/S$_0$ conical intersection. The XMCQDPT2/SA(3)-CASSCF(16,14)/(aug)-cc-pVDZ barrier of 0.41~eV is associated with reaching the lowest-lying S$_1$/S$_0$ CI with a sloped topography along the dark branch (see Fig.~\ref{fig:Pump_Probe_Diagram} and Fig.~S6). Quasi-equilibrium theory for a microcanonical ensemble predicts a statistical lifetime of 98~ps for the \1S excited state at 300~K after 400~nm excitation (Figs.~S7-S8), a value that agrees perfectly with our experimental findings.

\section{Conclusions}
The spectroscopy and excited-state dynamics of meta-HBDI were studied by performing action spectroscopy and femtosecond pump-probe experiments. Our action-photoabsorption spectroscopy studies in vacuum directly revealed the presence of an optically dark \1S state, originating from the $\pi-\pi^{*}$ transition with a CT character. We assessed the strong CT character of the \0S$\rightarrow$\1S transition by measuring the blue-shift of the \1S absorption band of meta-HBDI complexed with the betaine zwitterion. Femtosecond pump-probe experiments revealed $\sim$100~fs ultrafast relaxation out of the excited \2S state, followed by population and trapping in the \1S state for $\sim$100~ps. Our experimental observations were rationalized by high-level quantum-chemistry calculations. Three conical intersections were found that interconnect the \2S, \1S, and \0S states, which are relevant to the relaxation dynamics of the meta-chromophore in the gas phase. Although dark, the \1S state can be populated from the higher-lying bright \2S state by internal conversion that involves conical intersections. Importantly, this extremely efficient process out-competes and thus suppresses electron detachment from the \2S electronic resonance. The actual timescale for populating the dark state depends on the potential-energy topography, and the trapping lifetime is determined by energy barriers, along the minimum-energy pathway that leads from the planar \1S-minimum structure to the highly twisted \1S/\0S conical intersection. The charge-transfer character of the low-lying dark state provides a direct handle to control its lifetime via the polar environment. This offers a clear design principle for engineering synthetic chromophores with tailored photophysical properties.

Our gas-phase action-absorption and photoelectron spectroscopy study combined with high-level {\it ab initio} calculations has unequivocally identified a long-lived dark state that acts as an efficient trapping site in the meta-HBDI anion. This discovery and the detailed insight into the charge-transfer character establish a mechanistic framework for understanding photostability, in which dark states act as a general ``energy sink''. By dissipating excess electronic energy, these states suppress deleterious electron transfer and photodegradation following UV excitation. This principle, now revealed in the unperturbed chromophore, may extend to other environments, where similar states could be tuned to enhance photostability or, conversely, to design optical switches. Furthermore, it draws a compelling parallel to the function of dark states in carotenoids, which are critical for photoprotection. By providing direct, gas-phase spectroscopic access to these elusive states, our work establishes a new benchmark for unravelling complex excited-state landscapes, offering a fresh perspective on a cornerstone of biological photochemistry.

\section{Methods}
This study employs a unique combination of gas-phase photoabsorption detection, femtosecond time-resolved action spectroscopy, and photoelectron spectroscopy, supported by high-level quantum chemical calculations to interpret the results.

\subsection*{Action-absorption spectroscopy}
Action-photoabsorption spectroscopy measurements of meta-HBDI were performed at room temperature at the ion-storage ring ELISA.~\cite{nielsen2001,Andersen_2004} Methanol - dissolved chromophores were brought into the gas-phase by electrospray ionization, injected into the storage ring and photoexcited by a pulsed nanosecond-laser system (EKSPLA, NT232-50-SH-SFG). The number of neutral fragments after photoexcitation was used as a measure of the absorption cross-section after normalization to the number of photons and stored ions.

\subsection*{Pump-probe time-resolved action-absorption spectroscopy}
The relaxation dynamics was investigated by exciting the molecules by fs-laser pulses in the SAPHIRA ion-storage ring at Aarhus University \cite{Pedersen2015,Svendsen_2017,Kiefer_2019}. Here the molecules were excited with a 400~nm fs pump-laser pulse into \2S, and afterwards probed by a 800~nm fs-laser pulse at a variable pump-probe delay. In addition to the short time delay between the two fs-laser pulses, the molecular-response time by dissociation was recorded by the ion-storage technique. Thus, as pump-probe signal (as a function of the probe-pulse delay) we used the time-resolved yield of neutral photofragments in the storage ring. The advantage of the ion storage ring technique is that it enables the registration of prompt action (fragmentation within the first ~10~$\mu$s after photoexcitation) as well as delayed action. The counts of prompt neutral fragments after pump and (delayed) probe pulses were created largely by photodetachment rendering a neutral detectable product. The delayed action, on the other hand, appeared when the S$_2$ state returned to the S$_0$ ground state by internal conversion, where slow statistical fragmentation and thermionic electron emission occur, typically on the millisecond timescale.

\subsection*{Frequency- and time-resolved photoelectron spectroscopy}
Photoelectron spectroscopy measurements were carried out at Durham University\cite{Verlet2014_PES,Anstoter2016}. Briefly, electrons were detected and energy analyzed as a function of the photon energy to yield photoelectron spectra \cite{Lecointre2010}. In addition, the number of energy-selected photoelectrons were recorded as a function of pump-probe time delay to follow the excited-state relaxation dynamics.  

\subsection*{Ab initio calculations}
High-level {\it ab initio} calculations were performed using the multi-state multi-reference perturbation theory XMCQDPT2\cite{Granovsky2011} and complete active space self-consistent field method. The Firefly package~\cite{firefly} was used for all quantum chemistry calculations. The S$_1$ excited-state lifetime was computed as functions of ground-state temperature and excitation wavelength within the bright S$_0$ $\rightarrow$ S$_2$ transition using quasi-equilibrium theory for a microcanonical ensemble. The details are described in the Supporting Information.}\\

\newpage
\begin{acknowledgement}
Action-absorption studies were supported by Villum Fonden (Grant No. 17512, L.H.A.). Photoelectron studies were supported by the Engineering and Physical Sciences Research Council (Grant No. EP/Z53545X/1, J.R.R.V.). The theoretical part of this work was supported by the Russian Science Foundation (Grant No. 24-43-00041, A.V.B.). The calculations were carried out using the equipment of the shared research facilities of HPC computing resources at Lomonosov Moscow State University, as well as the local resources (RSC Tornado) provided through the Lomonosov Moscow State University Program of Development.
\end{acknowledgement}

\begin{suppinfo}
Supplementary information is available for this article, including computational details and additional computational results. 
\end{suppinfo}


\providecommand{\latin}[1]{#1}
\makeatletter
\providecommand{\doi}
  {\begingroup\let\do\@makeother\dospecials
  \catcode`\{=1 \catcode`\}=2 \doi@aux}
\providecommand{\doi@aux}[1]{\endgroup\texttt{#1}}
\makeatother
\providecommand*\mcitethebibliography{\thebibliography}
\csname @ifundefined\endcsname{endmcitethebibliography}  {\let\endmcitethebibliography\endthebibliography}{}

\end{document}